# A Novel Compressed Sensing Based Model for Reconstructing Sparse Signals Using Phase Sparse Character

Zhengli Xing[1], Jie Zhou[1*], Jiangfeng Ye[1], Jun Yan[1], Lin Zou[2], Qun Wan[2]
[1]Institute of Electronic Engineering, China Academy of Engineering Physics,
Mianyang, China
[2]School of Electronic Engineering, University of Electronics Science and Technology of China,
Chengdu, China
author_correspond@163.com

*Abstract*—**Phase modulation is a commonly used modulation mode in digital communication, which usually brings phase sparsity to digital signals. It is naturally to connect the sparsity with the newly emerged theory of compressed sensing (CS), which enables sub-Nyquist sampling of high-bandwidth to sparse signals. For the present, applications of CS theory in communication field mainly focus on spectrum sensing, sparse channel estimation etc. Few of current researches take the phase sparse character into consideration. In this paper, we establish the novel model of phase modulation signals based on phase sparsity, and introduce CS theory to the phase domain. According to CS theory, rather than the bandwidth, the sampling rate required here is scaling with the symbol rate, which is usually much lower than the Nyquist rate. In this paper, we provide analytical support for the model, and simulations verify its validity.**

## I. Introduction

As to digital modulation, phase, frequency and amplitude are commonly used parameters in digital communication. Using different parameters in the modulation procedure, there are various types of modulation, such as one-parameter modulation (PSK, ASK, FSK, OQPSK, MSK) and multi-parameter modulation (APSK, QAM). As the modulation parameters are different, different modulation types need different models to describe, resulting the variety of demodulation procedures, which also increases the complexity of the receiver.

On the other side, frequency resources are getting more and more scare as user increase sharply, which encourages the popularization of high-frequency and wideband signals. However, it undoubtedly poses a heavy burden to Analog to Digital Converter (ADC).

In recent years, Compressed Sensing (CS) theory has emerged, with the ability of sampling sparse high-bandwidth signals at sub-Nyquist rate. CS theory exploits the sparsity of high dimensional signals, and enables the reconstruction of signals from some nonuniform measurements. The number of required measurements depend on the sparsity of the signals, rather than the bandwidth. With the compressed measurements, it is possible to reconstruct the original signal by solving a convex optimization problem.

In this paper, based on the analysis of the phase modulation, we establish the sparse model of digital signals, whose constellation diagrams are sparse in the complex plane. As can be seen below, our model can be applied to all phase sparse signals, whose constellations are some discrete points, such as PSK, MSK, OQPSK, QAM, APSK, et al. Considering all the modulation parameters, the reconstruction can also be regarded as a joint demodulation procedure. In this model, the number of nonzero elements equals that of symbols. From CS theory, this means the rate of nonuniform Compressive Sampling (NCS) scales with the symbol rate. Whereas, according to the Nyquist sampling theorem, sampling rate should at least be two times of the bandwidth to avoid aliasing, which can be much higher than the symbol rate, especially for the wideband signals.

The remainder of this paper is organized as follows. In section II, we introduce the signal model and establish the sparse phase model. Section III describes the CS theory and the compressive sampling model. Discussions are made in Section IV, followed by simulation results in Section V. Finally, Section VI concludes the paper.

## II. Sparse Phase Model

For digital modulation, phase is a key parameter. And for many types of modulation utilizing phase, the constellation diagrams are some discrete points in the complex plane, showing sparse character.

### A. Signal Model

It is known that, the bit streams will be processed in order of mapping into constellation, square root raised cosine (SRRC) filtering and quadrature modulating.

Here, we define $\mathbf{r} \in \mathbb{C}^{N \times 1}$ to be the vector of uniform samples of the signal:

$$\mathbf{r} = [r_1, \cdots, r_N]^T \quad (1)$$

Taking the modulation procedure into consideration, the received signal can be decomposed as follows:

This research was funded in part by the National Nature Science Foundation of China under grant 61301267 and in part by the Fundamental Research Funds for the Central Universities.

$$\mathbf{r}_a = \begin{bmatrix} e_1 & & \\ & \ddots & \\ & & e_N \end{bmatrix} \begin{bmatrix} s_1 \\ \vdots \\ s_N \end{bmatrix} = \mathbf{Es} \quad (2)$$

and

$$\mathbf{E} = \begin{bmatrix} e_1 & & \\ & \ddots & \\ & & e_N \end{bmatrix} \in \mathbb{C}^{N \times N} \quad (3)$$

$$\mathbf{s} = [s_1, \cdots, s_N]^T \in \mathbb{C}^{N \times 1}$$

where $e_i = exp(j2\pi f_c t_i)$, and $f_c$ is the carrier frequency, $\mathbf{E}$ is the carrier wave matrix. $\mathbf{s}$ is the vector of symbols after baseband modulation. It can be seen that, (2) functions as the procedure of quadrature modulation.

If we denote the number of symbols after constellation as M. We can rewrite as follows:

$$\mathbf{s} = \mathbf{Fb} = \mathbf{F}_1\mathbf{Ub}$$
$$\mathbf{F} \in \mathbb{R}^{N \times M} \quad (4)$$
$$\mathbf{b} = [b_1, b_2, \cdots, b_M]^T \in \mathbb{C}^{M \times 1}$$

where $\mathbf{b}$ is the sequence of constellated symbols. $\mathbf{F}$ is the baseband shaping matrix, containing the procedure of SRRC filtering and interpolation. In order to express these two functions more clearly, we can decompose $\mathbf{F}$ into separate matrices, $\mathbf{F}_1$ and $\mathbf{U}$, which are filtering matrix and interpolation matrix respectively. And the expression of $\mathbf{F}_1$ and $\mathbf{U}$ are:

$$\mathbf{F}_1 = \begin{bmatrix} c_1 & \cdots & c_L & & & \\ & c_1 & \cdots & c_L & & \\ & & & \ddots & & \\ & & & & c_1 & \cdots & c_L \end{bmatrix} \in \mathbb{R}^{N \times N} \quad (5)$$

$$\mathbf{U} = \mathbf{I} \otimes \mathbf{y} \in \mathbb{R}^{N \times M}$$

and

$$\mathbf{c} = [c_1 \ \ldots \ c_L]$$

$$\mathbf{I} = \begin{bmatrix} 1 & & & \\ & 1 & & \\ & & \ddots & \\ & & & 1 \end{bmatrix} \in \mathbb{R}^{M \times M} \quad (6)$$

$$\mathbf{y} = [1 \ 0 \ \cdots \ 0]^T \in \mathbb{R}^{n_s \times 1}$$

where $\mathbf{c}$ is the coefficients vector of SRRC, $L$ is the number of coefficients. $L$ is decided by the order of SRRC, and $\mathbf{c}$ is decided by the order and the roll-off factor $\alpha$ of SRRC. $\mathbf{I}$ is an identity matrix, whose order equals the number of symbol number. $\mathbf{y}$ is the interpolation vector, $M$ is the symbol number, $n_s = N/M$ is the rate of interpolation, '$\otimes$' denotes the Kronecker product.

As to $\mathbf{b}$, different types of signals have different forms.

For PSK, the constellation model is []:

$$b_k = \exp(j2\pi(m_k - 1)/M') \quad (7)$$

where $M' \in \{2, 4, 8 \cdots\}$ is the number of unique phases used, $m_k$ is the $k$th transmitted symbol.

For rectangular QAM, the signal model is [00987656]:

$$b_k = A_I + jA_Q \quad (8)$$

where $A_I$ and $A_Q$ are the signal amplitudes of in-phase and quadrature components, respectively. In an $I \times Q$ R-QAM scheme, $log_2(I \cdot Q)$ bits of serial information stream are grouped to be mapped onto 2-dimensional signal constellation. Among the grouped information, $log_2 I$ bits are mapped onto the in-phase channel, whose amplitude $A_I$ is selected over the set of $\{\pm d, \pm 3d, \cdots, \pm(I-1)d\}$. Similarly, the $log_2 Q$ bits are mapped onto the quadrature channel, whose amplitude $A_Q$ is selected over the set of $\{\pm d, \pm 3d, \cdots, \pm(Q-1)d\}$.

For APSK, the signal constellation point $b_k$ is drawn from a set $\mathbf{X}$ given by []:

$$\mathbf{X} = \begin{cases} r_1 e^{j\left(\frac{2\pi}{n_1}i + \theta_1\right)} & i = 0, \ldots, n_1 - 1 \\ r_2 e^{j\left(\frac{2\pi}{n_2}i + \theta_2\right)} & i = 0, \ldots, n_2 - 1 \\ \vdots & \\ r_{n_R} e^{j\left(\frac{2\pi}{n_R}i + \theta_{n_R}\right)} & i = 0, \ldots, n_{n_R} - 1 \end{cases} \quad (9)$$

where we have defined $n_l$, $r_l$ and $\theta_l$ as the number of points, the redius and the relative phase shift corresponding to the $l$-th ring respectively. We can nickname such modulations as $n_1 + n_2 + \cdots + n_{n_R} - APSK$.

The constellation diagram for QPSK, 16QAM and 4-12APSK are shown in Figure 1(a)-(c) respectively.

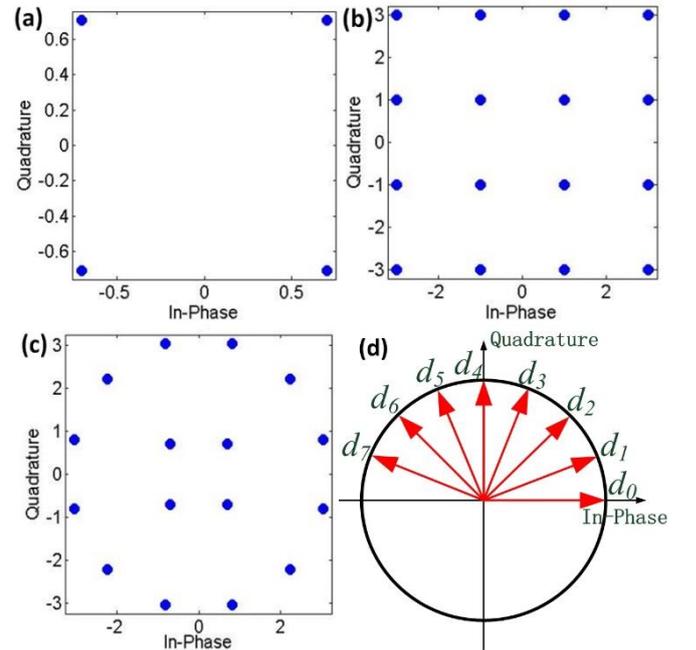

Figure. 1. Constellation diagrams for QPSK, 16QAM, 4-12APSK and the sparse vector

Hence, $\mathbf{r}$ can be rewritten as:

$$\mathbf{r} = \mathbf{EF}_1\mathbf{Ub} \quad (10)$$

*B. Sparse Phase model*

From the signal model defined in Section II-A, we can see that, for MPSK, MQAM and $n_1 + n_2 + \cdots + n_{n_R} - APSK$ ($M = n_1 + n_2 + \cdots + n_{n_R}$) signals, the modulated symbol can only vary within $M$ possibilities, where $M$ is the order of modulation. Thus, the constellated sequence **b** can be decomposed further.

Such as QPSK, **b** can be rewritten as:

$$\mathbf{b} = \begin{bmatrix} e^{j\frac{\pi}{4}} & e^{j\frac{3\pi}{4}} & e^{j\frac{5\pi}{4}} & e^{j\frac{7\pi}{4}} & 0 & 0 & \cdots & & \\ 0 & \cdots & & e^{j\frac{\pi}{4}} & e^{j\frac{3\pi}{4}} & e^{j\frac{5\pi}{4}} & e^{j\frac{3\pi}{2}} & 0 \\ & & & & & & & \ddots \\ 0 & & \cdots & & & & 0 & e^{j\frac{\pi}{4}} \end{bmatrix}$$

$$= \mathbf{Q}_{QPSK} \mathbf{b}^*_{QPSK} \quad (11)$$

and

$$b_k = \begin{bmatrix} e^{j\frac{\pi}{4}} & e^{j\frac{3\pi}{4}} & e^{j\frac{5\pi}{4}} & e^{j\frac{7\pi}{4}} \end{bmatrix} \begin{bmatrix} b_{k,1} \\ b_{k,2} \\ b_{k,3} \\ b_{k,4} \end{bmatrix} \quad (12)$$

$$= e \cdot (b^*_{QPSK,k})^T$$

where '·' denotes the inner product, and

$$e = \begin{bmatrix} e^{j\frac{\pi}{4}} & e^{j\frac{3\pi}{4}} & e^{j\frac{5\pi}{4}} & e^{j\frac{7\pi}{4}} \end{bmatrix} \quad (13)$$

$$b^*_{QPSK,k} = \begin{bmatrix} b_{k,1} & b_{k,2} & b_{k,3} & b_{k,4} \end{bmatrix}^T$$

For 16QAM, b can be rewritten as:

$$\mathbf{b} = \begin{bmatrix} A_1 & \cdots & A_{16} & & & & \\ & & & A_1 & \cdots & A_{16} & & \\ & & & & & & \ddots & \\ & & & & & & A_1 & \cdots & A_{16} \end{bmatrix} \begin{bmatrix} b_{1,1} \\ \vdots \\ b_{1,16} \\ \vdots \\ b_{M,1} \\ \vdots \\ b_{M,16} \end{bmatrix} \quad (14)$$

$$= \mathbf{Q}_{16QAM} \mathbf{b}^*_{16QAM}$$

And for $n_1 + n_2 + \cdots + n_{n_R} - APSK$, there is similar conclusion.

It is obvious that, there is only one non-zero component of $b^*_{QPSK,k}$, and the value is 1. This conclusion also works for 16QAM and $n_1 + n_2 + \cdots + n_{n_R} - APSK$.

From the discussion above, we can see that, for the phase sparse modulation types, PSK, QAM and APSK, the constellated sequence has the following decomposition:

$$\mathbf{b} = \mathbf{Q}\mathbf{b}^* \quad (15)$$

However, for different modulation types, the matrix **Q** and vector $\mathbf{b}^*$ are different. In order to get a unified expression, we define the sparse vector **d**, which divides the unit semicircle into $J$ equal parts. For simplicity, we set $J = 8$, as shown in (17) and Figure 1(d).

From Figure 1 and the constellation model (7)-(9), it can be seen that, the constellation points of PSK, QAM and APSK signals are sparse in the complex plane. Here, we purpose a new constellation expression:

$$\mathbf{b} = \mathbf{D}\mathbf{\theta} \quad (16)$$

and

$$\mathbf{D} = \begin{bmatrix} d_1 & \cdots & d_J & 0 & & \cdots & \\ 0 & \cdots & 0 & d_1 & \cdots & d_J & 0 & \cdots \\ & & & & & & \ddots & \\ 0 & & & & & 0 & d_1 & \cdots & d_J \end{bmatrix}$$

$$= \mathbf{I} \otimes \mathbf{d} \in \mathbb{C}^{M \times JM}$$

$$\mathbf{d} = \begin{bmatrix} d_1 & \cdots & d_J \end{bmatrix} \quad (17)$$

$$d_i = \exp\left(j\frac{\pi(l-1)}{J}\right)(l = 1, 2, \cdots, J)$$

$$\mathbf{\theta} = [\theta_{1,1}, \cdots, \theta_{1,J}, \cdots, \theta_{k,1}, \cdots, \theta_{k,J}, \cdots]^T \in \mathbb{R}^{J \cdot M \times 1}$$

According to (16) and (17), the $k$th element of **b** is calculated as follows:

$$b_k = \mathbf{d} \cdot \mathbf{\theta}_k^T \quad (18)$$

where $\mathbf{\theta} = [\theta_{k,1}, \cdots, \theta_{k,J}]^T$, '·' denotes the inner product.

For MPSK ($M$=2, 4, 8), there only exists one non-zero element of $\mathbf{\theta}_k$ and its value is 1. As the symbols are randomly chosen from the symbol set, the location of non-zero element is also random. Therefore, the number of non-zero elements in $\mathbf{\theta}$ is $M$, while the length of $\mathbf{\theta}$ is $J \cdot M$, which means $\mathbf{\theta}$ is sparse.

However, for QAM and APSK signals, the situations may be a little different. Seeing from the constellation diagram, it is clear that, for those constellation points being identical with MPSK constellations, same conclusion can be drawn. But for other points, they can be expressed by the linear combination of two vectors of **d**, which means there are two non-zero elements of $\mathbf{\theta}_k$, and their value may also be not 1. Thus, for QAM and APSK, the number of non-zero elements of $\mathbf{\theta}$ is more than that of MPSK, but we can safely claim that the times are in the range of (1, 2), and the non-zero elements in $\mathbf{\theta}_k$ ranges between $M$ and $2M$. Therefore, $\mathbf{\theta}$ is also sparse for QAM and APSK signals.

Considering all the discussion above in this section, **r** can be rewritten as:

$$\mathbf{r} = \mathbf{EF_I UD\theta} = \mathbf{\Psi\theta} \quad (19)$$

where $\mathbf{\Psi} = \mathbf{EF_I UD}$.

If we regard the matrix $\mathbf{\Psi}$ as a basis, then we can get the conclusion that, **r** is sparse on the basis, and the sparsity concerns with the signal type and the parameter $J$. Now, we've established the phase sparse model, and it is natural to connect it with the CS theory.

III. COMPRESSED SENSING OF PHASE SPARSE SIGNALS

*A. Compressed Sensing*

In CS theory, if the measured signal **x** is sparse on a basis $\mathbf{\Omega}$:

$$\mathbf{x} = \mathbf{\Omega\beta} \quad (20)$$

and there are only $k$ coefficients of $\mathbf{\beta}$ much larger than zero.

**x** is called sparse or compressible, if $k \ll N'$, where $N'$ is the length of **x**. It has been proven that x can be reconstructed form L non-adaptive linear projection measurements, where $L = kO(logN')$ [].

Let $y = \mathbf{\Phi x} = \mathbf{\Phi\Omega\beta}$, and $\mathbf{\Phi} \in \mathbb{R}^{L \times N}$ is the measurement

matrix. The reconstruction of **x** converts to the following optimization problem []:

$$\min \|\boldsymbol{\beta}\|_1 \quad s.t. \quad \mathbf{y} = \boldsymbol{\Phi}\mathbf{x} \tag{21}$$

where $\|\cdot\|_1$ stands for the $l_1$ norm.

### B. Signal Reconstruction based on Compressive Sampling

In CS theory, random matrices, e.g., Gaussian matrix and Bernoulli matrix, are commonly used measurement matrices. Here, in this paper, we can design the measurement matrix as Gaussian random matrix, whose coherence with a fixed orthonormal basis is very low [].

According to (19) and (21), the problem of recovering phase sparse signals from compressive measurements is a convex problem as follows:

$$\begin{aligned}\hat{\boldsymbol{\theta}} &= \arg\min \|\boldsymbol{\theta}\|_1 \\ s.t. \quad &\|\mathbf{y} - \boldsymbol{\Phi}\mathbf{r}\|_2 \leq \varepsilon \\ &\mathbf{r} = \boldsymbol{\Psi}\boldsymbol{\theta}\end{aligned} \tag{22}$$

Here, we solve the convex problems by CVX [].

Then, we get the constellation sequence $\hat{\mathbf{b}} = \mathbf{D}\hat{\boldsymbol{\theta}}$, and the transmitted signal $\mathbf{r} = \boldsymbol{\Psi}\boldsymbol{\theta}$.

## IV. DISCUSSION

### A. Amendment for the Sparse Phase Model

Now, we've established the sparse phase model for most phase sparse signals, such as PSK, QAM and APSK discussed above, this model also works for other signals, whose constellation diagrams are some discrete points in the complex plane, e.g. ASK, MSK and OQPSK. In fact, we have to point out that for some types of modulation, the model needs a little amendment. Here, we just discuss OQPSK for example.

For OQPSK, the signal model is as follows:

$$\begin{aligned}s_{OQPSK,an} &= A[I_n + j*Q_n]\exp(j2\pi f_c t) \\ I_n &= \sum_n a_n g(t - 2nT_b) \\ Q_n &= \sum_n b_n g(t - (2n+1)T_b)\end{aligned} \tag{23}$$

where $a_n, b_n \in \{\pm 1\}$ are i.i.d. (independently identically distributed) random sequences. $T_b$ is the bit period.

From the signal model above, it can be seen that, the only difference between OQPSK and QPSK is that the in-phase components have a delay of half symbol period relative to the quadrature components. Thus, we have to decompose the matrices $\mathbf{D}$, $\mathbf{U}$, $\mathbf{F}_1$ into in-phase part and quadrature part:

$$\overline{\mathbf{D}} = \begin{bmatrix} \mathbf{D}_R & \mathbf{Z}_D \\ \mathbf{Z}_D & \mathbf{D}_I \end{bmatrix} \in \mathbb{R}^{2M \times 2JM} \tag{24}$$

$$\mathbf{D}_R = real(\mathbf{D}) \qquad \mathbf{D}_I = imag(\mathbf{D})$$

where $\mathbf{Z}_D$ is a zero matrix of the same size of $\mathbf{D}$.

$$\begin{aligned}\overline{\boldsymbol{\theta}} &= [\boldsymbol{\theta}_R \ \vdots \ \boldsymbol{\theta}_I]^T \in \mathbb{R}^{2JM \times 1} \\ \boldsymbol{\theta}_R &= real(\boldsymbol{\theta}) \qquad \boldsymbol{\theta}_I = imag(\boldsymbol{\theta})\end{aligned} \tag{25}$$

$$\overline{\mathbf{U}} = \begin{bmatrix} \mathbf{U} & \mathbf{Z}_U \\ \mathbf{Z}_U & \mathbf{U} \end{bmatrix} \in \mathbb{R}^{2N \times 2M} \tag{26}$$

where $\mathbf{Z}_U$ is a zero matrix of the same size of $\mathbf{U}$.

$$\overline{\mathbf{F}_1} = \begin{bmatrix} \mathbf{F}_1 & \mathbf{Z}_{F_1} \\ \mathbf{Z}_{F_1} & \mathbf{F}_1 \end{bmatrix} \in \mathbb{R}^{2N \times 2N} \tag{27}$$

where $\mathbf{Z}_{F_1}$ is a zero matrix of the same size of $\mathbf{F}_1$.

As to the matrix $\mathbf{E}$, there is no need to change.

What's more, in order to fulfil the relative delay between in-phase components and quadrature components, a delay matrix $\mathbf{P}$ needs to be added between $\mathbf{E}$ and $\mathbf{F}_1$:

$$\mathbf{P} = \begin{bmatrix} & & & * & & & & & & \\ 0 & \cdots & 0 & 1 & 0 & \cdots & j & 0 & \cdots & 0 \\ & \cdots & & 0 & 1 & \cdots & & j & 0 & \cdots \\ & & & & & \ddots & & & \ddots & \end{bmatrix} \in \mathbb{C}^{N \times 2N} \tag{28}$$

where the column with '*' is the $n_s/2+1$-th column.

We can that, the delay matrix realizes the function of half symbols' delay by a deliberately mismatch of in-phase and quadrature components.

Thus, the phase sparse model for OQPSK is:

$$\mathbf{r}_{a,OQPSK} = \mathbf{E}\mathbf{P}\overline{\mathbf{F}_1}\overline{\mathbf{U}}\overline{\mathbf{D}}\overline{\boldsymbol{\theta}} = \overline{\boldsymbol{\Psi}}\overline{\boldsymbol{\theta}} \tag{29}$$

where

$$\overline{\boldsymbol{\Psi}} = \mathbf{P}\overline{\mathbf{F}_1}\overline{\mathbf{U}}\overline{\mathbf{D}} \tag{30}$$

The CS model of OQPSK signals based on the phase sparse model is as follows:

$$\begin{aligned}\overline{\boldsymbol{\theta}} &= \arg\min \|\overline{\boldsymbol{\theta}}\|_1 \\ s.t. \quad &\|\mathbf{y} - \boldsymbol{\Phi}\mathbf{r}\|_2 \leq \varepsilon \\ &\mathbf{r} = \overline{\boldsymbol{\Psi}}\overline{\boldsymbol{\theta}}\end{aligned} \tag{31}$$

In fact, without using of delay matrix $\mathbf{P}$, this model also works for the signals of PSK, QAM and APSK.

For MSK, the signal model is [06624684]:

$$\begin{aligned}s_{MSK,an} &= [I_n + j*Q_n]\exp(j\omega_0 t) \\ I_n &= \sum_n a_n rect(t-(2n-1)T_b)cos(\pi t/2T_b) \\ Q_n &= \sum_n b_n rect(t-2nT_b)sin(\pi t/2T_b)\end{aligned} \tag{32}$$

where $a_n, b_n \in \{\pm 1\}$ are i.i.d. (independently identically distributed) random sequences. $T_b$ is the bit period.

We can see that MSK signal model is similar to that of OQPSK, except for the shaping filter is different. So the difference of phase sparse model between MSK and OQPSK is the coefficients of filtering matrix $\mathbf{F}_1$.

Due to the limited space here, we can't discuss all the signals. And for all the phase sparse signals, the model should

### B. Sparse Phase Model for Multi-Signals

All the discussions above deal with the single signal situation, when it comes to the multi-signals situation, the model should be as follows:

$$\begin{aligned}\min \ &\sum_{i=1}^{K} \|\boldsymbol{\theta}_i\|_1 \\ s.t. \quad &\|\mathbf{y} - \boldsymbol{\Phi}\mathbf{r}\|_2 \leq \varepsilon \\ &\mathbf{r} = \sum_{i=1}^{i=K} \mathbf{r}_i \qquad \mathbf{r}_i = \boldsymbol{\Psi}_i \boldsymbol{\theta}_i\end{aligned} \tag{33}$$

where $K$ is the signal number, $\mathbf{r}_i$ is the $i$th signal, $\mathbf{\Psi}_i$ is the sparse basis for the $i$th signal, $\mathbf{\theta}_i$ is the sparse vector of $\mathbf{r}_i$ on the basis of $\mathbf{\Psi}_i$.

As to multi-signals model, the measured signal is the mixture of several signals. And shown in (33), different signal may have different sparse matrices.

### C. Discussion about the Measurement Number

In CS theory, one important parameter is the sparsity. Here, we define the sparse ratio as the ratio of non-zero components' number to the total number:

$$\lambda_\theta = \|\mathbf{\theta}\|_0 / |\mathbf{\theta}|_c \qquad (34)$$

where $\|\cdot\|_0$ is the $l_0$ norm, which measures the number of non-zero components [1]. $|\cdot|_c$ is the cardinality.

Based on the discussion of Section II and III, it is easy to conclude:

$$\begin{cases} \lambda_\theta = 1/J & \text{for PSK, OQPSK and MSK} \\ 1/J < \lambda_\theta < 2/J & \text{for QAM and APSK} \end{cases} \qquad (35)$$

For the Gaussian measurement matrix we use here, the number of measurements needed is [1]:

$$K \sim O\left(M \log \frac{1}{\lambda_\theta}\right) \sim O(M \log J) \qquad (36)$$

In CS theory, one key concept is so-call 'Information', and in the phase sparse model proposed here, it is clear that the 'Information' is in fact the symbols. According to (36), the compressive sampling rate scales with the symbol rate $R_s$, and the ratio is $c \cdot \log J$ ($c$ is a constant ranging in [1, 2)). Whereas, from the Nyquist sampling rule, the sampling rate should be more than 2 times of the bandwidth, which usually scales with the highest frequency of the signal and much higher than the symbol rate. That means, compressive sampling can greatly reduce the sampling rate.

### V. SIMULATIONS

In this section, we give some simulation results about model we proposed. Simulation scenarios are set as follows: $M = 64$, roll-off factor of SRRC $\alpha = 0.35$, $f_c = 400Hz$, $R_s = 100HZ$. Here, the Nyquist rate of uniform sampling is $f_s = 1.6kHz$ to avoid aliasing, which means $N = 1024$.

We can define the compression ratio as:

$$\eta = K/L \qquad (37)$$

In the simulation below, we set $K = 3M = 192$, i.e. $\eta = 0.1875$. Figure 2(a) depicts the constellation diagram of QPSK signal, and Figure 2(b) is the reconstruction result. Figure 2(c) is the simulation result of $\mathbf{\theta}$ and Figure 2(d) presents the partition amplification result. Since we set $J = 8$ in our simulation, Figure 2(d) shows that, there is only one non-zero element in every 8 elements of the solution for QPSK signal, which is consistent with our model.

Figure 2(e) and (f) are the real part and imaginary part of the original transmitted signal and the reconstructed results separately. It is obvious that, the reconstruction results are identical with the original data.

Figure 3(a)-(e) and Figure 4(a)-(e) are the simulation results of 16QAM and 16APSK respectively. As shown in Figure 3(c) and Figure 4(c), since the constellation points of QAM and APSK and more than one amplitudes, the solutions also have several amplitudes. Whereas, solutions are sparse.

Figure 5(a)-(e) are the corresponding simulation results of OQPSK signal, we can see the results are consistent with our analysis in Section IV-A.

Figure 6(a)-(f) are the simulation results of multi-signals. In the simulation, the mixed signal consists of two independent signal, QPSK and 16QAM. The simulation For the QPSK signal, the scenarios are set as follows: For QPSK signal, $M = 64$, $\alpha = 0.35$, $f_c = 400Hz$, $R_s = 100Hz$, $f_s = 1.6kHz$. For 16QAM, $M = 64$, $\alpha = 0.35$, $f_c = 500Hz$, $R_s = 200Hz$, $f_s = 3.2kHz$.

Figure 6(a) and Figure 6(b) are the reconstructed QPSK and 16QAM signals' constellation diagrams, we can see the reconstructed constellation diagrams are in accordance with the theoretical ones. Figure 6(c) shows the mixed signal's constellation diagram, in which QPSK and 16QAM signals' constellations have no influence on each other. Figure 6(d) and Figure 6(e) are the sparse vectors. As discussed above, for QPSK signal, the sparse vector has only one amplitude, and for 16QAM signal, the sparse vector has more than one amplitudes.

Figure 7(a)-(d) are the simulation results of the constellations of QPSK signals, with carrier phase deviation, timing offset, carrier frequency offset, and Gaussian white noise separately. As to Figure 7(a), we set the carrier phase deviation as $\pi/4$ in the simulation. It can be seen that, the residual carrier phase is transferred to the constellation diagram, resulting the $\pi/4$ full rotation of the constellation. Similar analyses can be made to Figure 7(b)-(d).

### VI. CONCLUSION

Based on the analysis of digital signal's modulation procedure, we proposed the phase sparse model. Introducing the CS theory, compressed sampling is applied to phase sparse signals, which can greatly reduce the required sampling rate. Relative simulation verified the proposed model.

What we have to admit is that, some issues, such as how to obtain the exact carrier frequency, carrier phase deviation, and how to realize symbol timing and carrier frequency synchronization before applying our model still remain to be further studied in our future work.

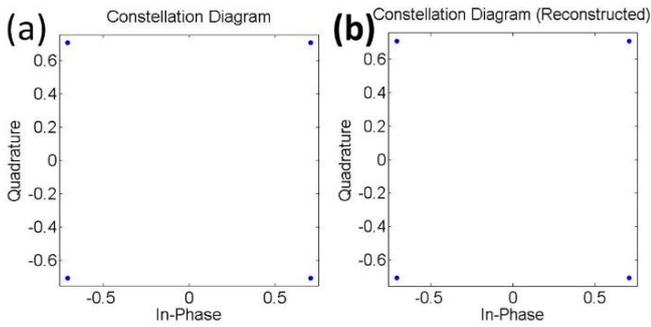
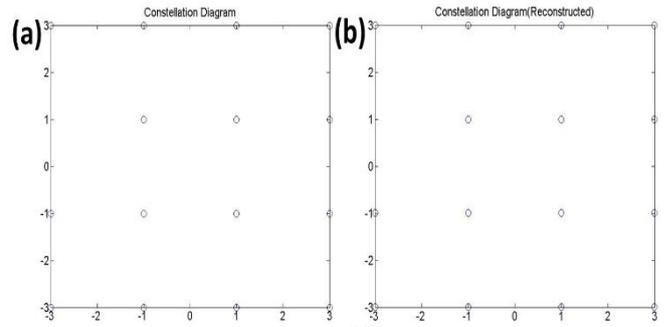
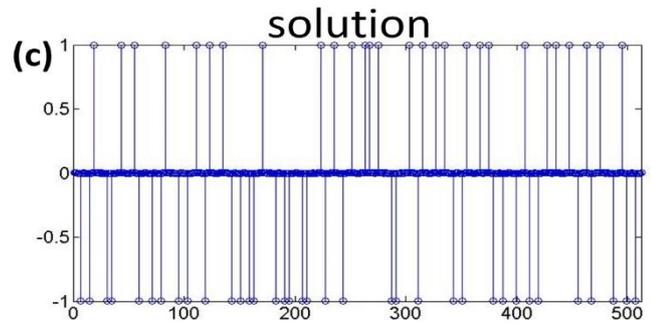
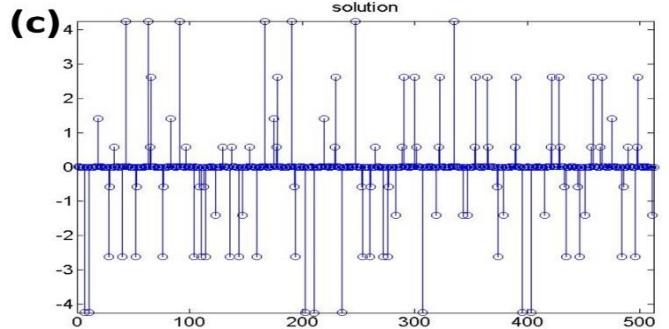
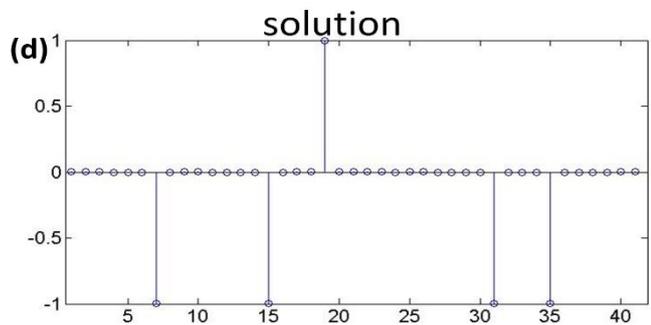
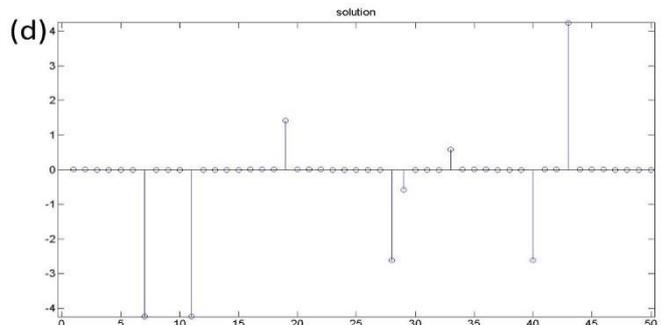
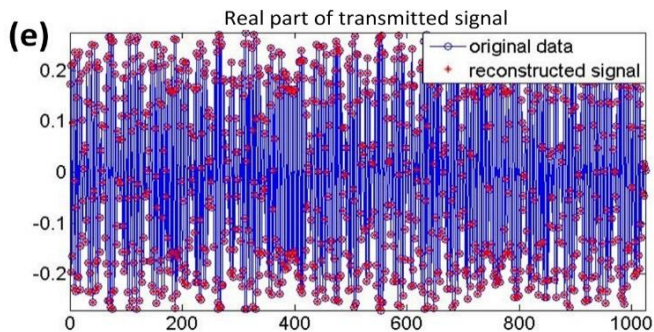
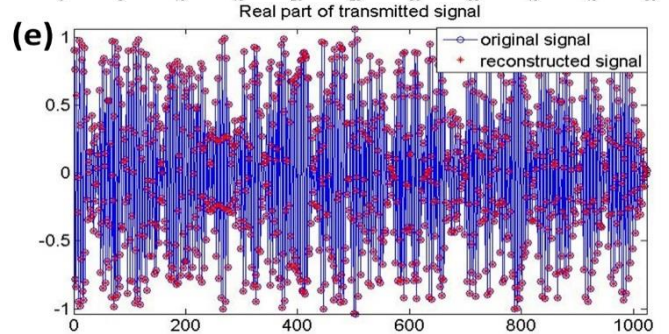
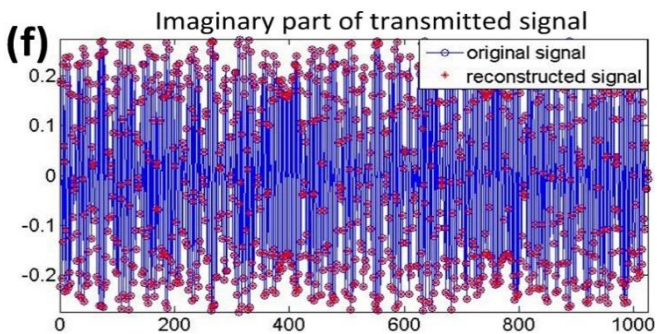
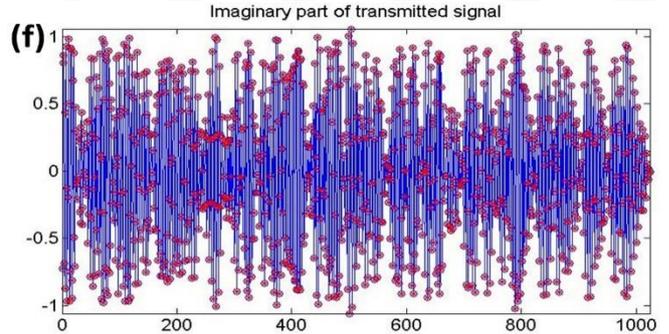

Figure. 2.  Simulation results of QPSK signals

Figure. 3.  Simulation results of 16QAM signals

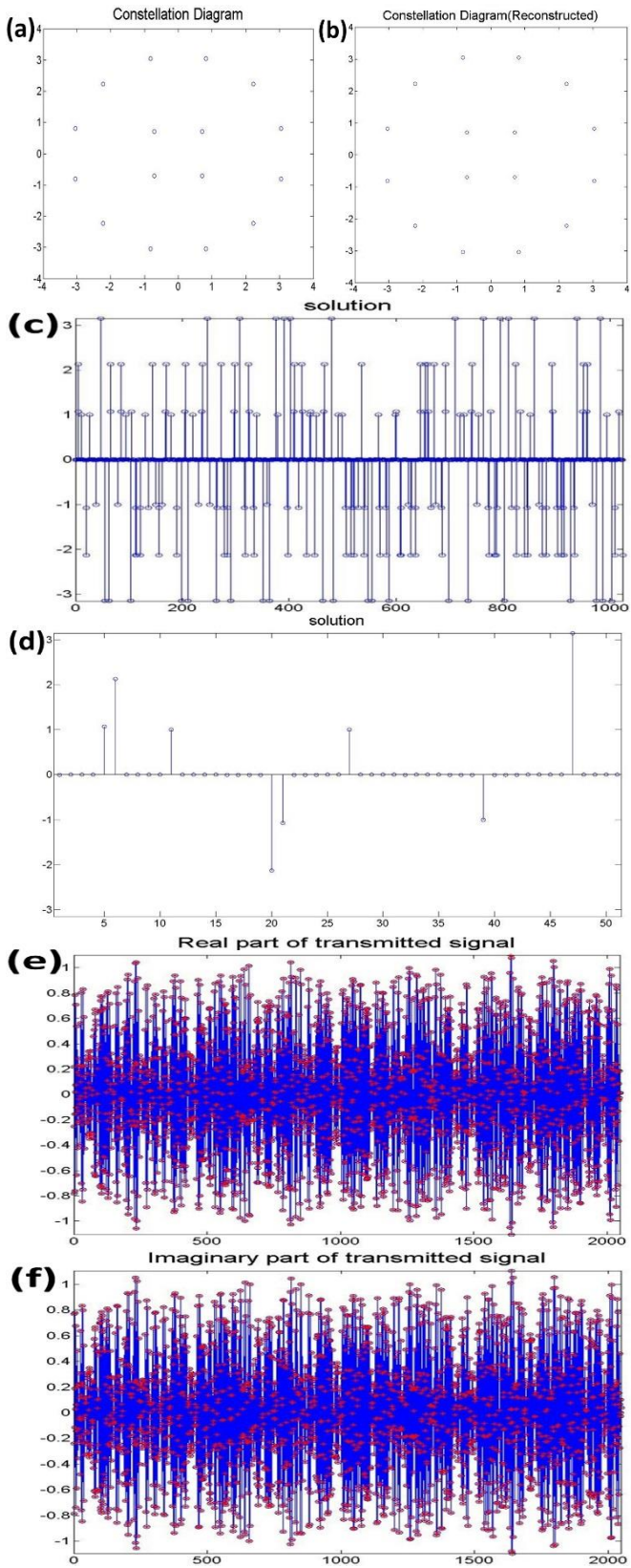

Figure. 4. Simulation results of 4-12APSK signals

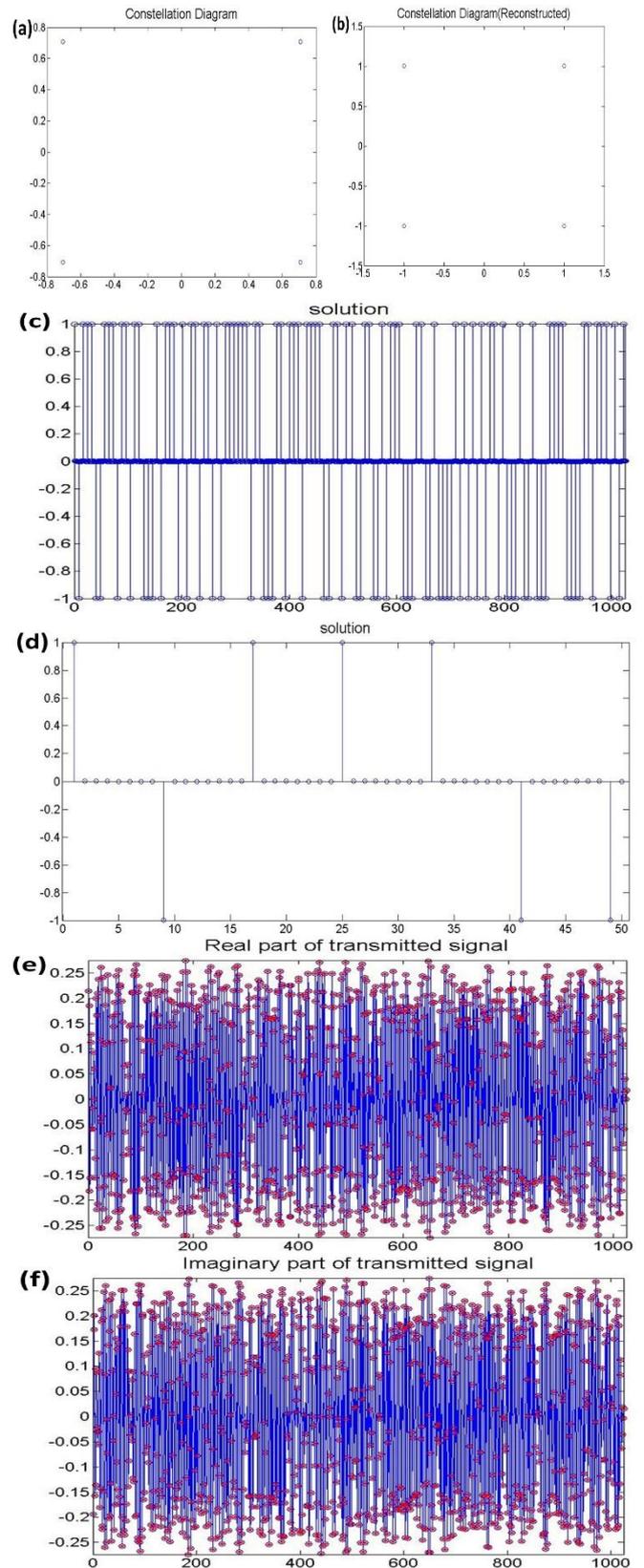

Figure. 5. Simulation results of OQPSK signals

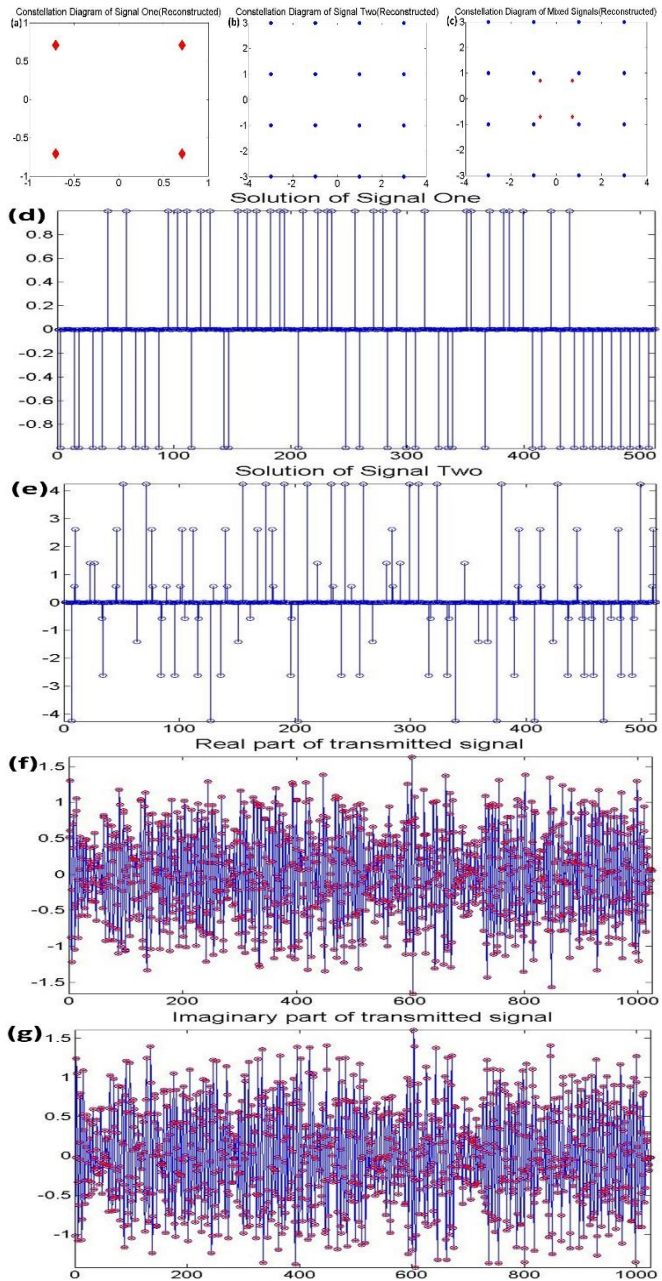

Figure. 6. Simulation results of mixed signals

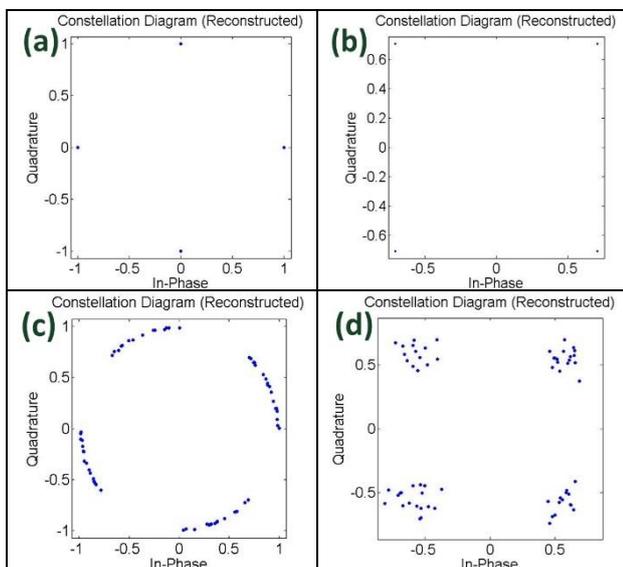

Figure. 7. Simulations with carrier phase deviation, timing offset, carrier frequency offset, and Gaussian white noise